\documentclass[twocolumn,showpacs,superscriptaddress,preprintnumbers,amsmath,amssymb,prb]{revtex4}

\usepackage{graphicx}
\usepackage{dcolumn}
\usepackage{bm}
\usepackage{amssymb}
\usepackage{color,soul}

\begin{document}

\title{Optically induced Aharonov-Bohm effect in mesoscopic rings}

\author{H. Sigurdsson}
\affiliation{Division of Physics and Applied Physics, Nanyang
Technological University 637371, Singapore} \affiliation{Science
Institute, University of Iceland, Dunhagi-3, IS-107, Reykjavik,
Iceland}

\author{O.V. Kibis}\email{Oleg.Kibis@nstu.ru}
\affiliation{Department of Applied and Theoretical Physics,
Novosibirsk State Technical University, Karl Marx Avenue 20,
Novosibirsk 630073, Russia} \affiliation{Division of Physics and
Applied Physics, Nanyang Technological University 637371,
Singapore}

\author{I.A. Shelykh}
\affiliation{Division of Physics and Applied Physics, Nanyang
Technological University 637371, Singapore} \affiliation{Science
Institute, University of Iceland, Dunhagi-3, IS-107, Reykjavik,
Iceland}


\begin{abstract}
We show theoretically that strong electron coupling to circularly
polarized photons in non-singly-connected nanostructures results
in the appearance of an artificial gauge field that changes the
electron phase. The effect arises from the breaking of
time-reversal symmetry and is analogous to the well-known
Aharonov-Bohm phase effect. It can manifest itself in the
oscillations of conductance as a function of the intensity and
frequency of the illumination. The theory of the effect is
elaborated for mesoscopic rings in both ballistic and diffusive
regimes.
\end{abstract}


\maketitle

\section{Introduction} Progress in modern nanotechnologies has
resulted in rapid developments in the fabrication of mesoscopic
objects, including non-singly-connected nanostructures such as
quantum rings. The fundamental interest attracted by these systems
is caused by a wide variety of purely quantum-mechanical
topological effects which can be observed in ring-like mesoscopic
structures. The most notable phenomenon amongst them is the
Aharonov-Bohm (AB) effect arisen from the direct influence of a
vector potential on the electron phase. \cite{AharonovBohm,
Chambers1960} In the ballistic regime, this effect results in
magnetic-flux-dependent oscillations of the conductance in
ring-like structures with a period equal to the fundamental
magnetic flux quantum $\Phi_0=h/|e|$. \cite{Webb,Timp,Vegvar,Wees}
In the diffusive regime, a second type of conductance oscillations
with the period $\Phi_0/2$ can be observed. They are known as the
Altshuler-Aronov-Spivak (AAS) oscillations and are associated with
the weak localization of electrons.
\cite{AAS,Sharvin1981,Bergmann1983, Pannetier1985}

From a fundamental viewpoint, the AB-AAS oscillations arise from
the broken time-reversal symmetry in the electron system
(conducting mesoscopic ring)  subjected to a magnetic flux through
the ring. Namely, the flux breaks the equivalence of clockwise and
counterclockwise electron rotations inside the ring, which results
in the flux-controlled interference of electron waves
corresponding to these rotations. The similar broken equivalence
of electron motion for mutually opposite directions caused by a
magnetic field can take place in various nanostructures, including
quantum wells, \cite{Kibis1998} carbon nanotubes \cite{Kibis2002}
and hybrid semiconductor/ferromagnet nanostructures.
\cite{Kibis2002_1} However, the time-reversal symmetry can be
broken not only by a magnetic flux but also by a circularly
polarized electromagnetic field. Indeed, the field breaks the
symmetry since time reversal turns clockwise polarized photons
into counterclockwise polarized ones and vice versa. As a result,
the electron coupling to circularly polarized photons can change
electron energy spectrum of quantum rings. \cite{Kibis2011}
Therefore, phenomena similar to the AB effect can take place in
ring-like electronic systems interacting with a circularly
polarized electromagnetic field. We will show below that the
conductance of these electron-photon systems can exhibit
oscillations which are formally equivalent to the AB-AAS
oscillations induced by a magnetic flux. The phenomenon can be
described in terms of an artificial $U(1)$ gauge field generated
by the strong coupling between electron and circularly polarized
photons. The theory of such optically-induced AB effect, which
lies at the border between condensed-matter physics and quantum
optics, is developed in this paper.

The paper is organized as follows. In the Section II, we introduce
the Schr\"{o}dinger problem describing the electron interaction
with circularly polarized photons in mesoscopic rings. The Section
III is devoted to derivation of an artificial $U(1)$ gauge field
arisen from the strong electron-photon coupling in the rings. In
the Section IV, AB-AAS oscillations of conductance caused by the
gauge field are analyzed.

\section{The model} 
\begin{figure}[th]
\includegraphics[width=0.48 \textwidth]{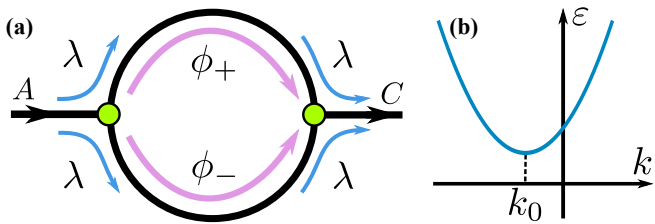}
\caption{(Color online) \textbf{(a)} The scheme of the electron
interference device consisting of an one-dimensional mesoscopic
ring which is connected with two one-dimensional leads at the
quantum point contacts (QPCs). An electron wave, which enters into
the device with the amplitude $A=1$, is split between the two
different paths with the transmission amplitudes $\lambda$ and
exits the device with the amplitude $C$.  The phase shift of the
electron waves traveling clockwise and counterclockwise inside the
ring, $\Delta\phi= \phi_+-\phi_-$, arises from the coupling to an
external electromagnetic field; \textbf{(b)} The scheme of the
electron energy spectrum $\varepsilon(k)$ in a mesoscopic ring
subjected to a circularly polarized electromagnetic field or a
stationary magnetic field. The spectrum is shifted along the $k$
axis by the wave vector $k_0$ which depends on the parameters of
the field. } \label{fig1}
\end{figure}

Let us consider the conventional model of
an electron interference device (see, e.g., Refs.
\cite{Gefen1983,Buttiker1984,Shelykh2005}) consisting of an
one-dimensional mesoscopic ring with radius $R$ and two
one-dimensional leads which are connected at the quantum point
contacts (see Fig.~\ref{fig1}a).
Generally, the phase shift between the clockwise and
counterclockwise traveling electron waves,
\begin{equation}\label{tau}
\Delta\phi=\phi_+ - \phi_-,
\end{equation}
can be nonzero: The shift can be caused by the application of an
external magnetic field (AB effect) or result from spin-orbit
interaction.
\cite{Aronov1993,Morpurgo1998,Frustaglia2004,Bergsten2006}
Experimentally, it can be detected by measuring the
field-dependent oscillations of the conductance of the device.

In order to write the phase shift (\ref{tau}) as a function of the
field parameters, we have to consider the electron energy spectrum
of an isolated ring subjected to an electromagnetic field with the
vector potential $\mathbf{A}$. If the field is time-independent,
then the electron energy spectrum can be found from the stationary
Schr\"odinger equation with the Hamiltonian
\begin{equation}\label{H0}
{\hat H}_0=\frac{1}{2m_e}\left({{\hat p}_\varphi}-eA_\varphi
\right)^2,
\end{equation}
where $\varphi$ is the electron angular coordinate in the ring,
${\hat p}_\varphi=-i(\hbar/R)\partial/\partial\varphi$ is the
operator of electron momentum in the ring, $e$ is the electron
charge, and $m_e$ is the effective electron mass in the ring.
Particularly, in the well-known case of a stationary magnetic
field, $\mathbf{B}$, directed perpendicularly to the ring plane,
the electron energy spectrum of the ring has the form
$$\varepsilon(m)=\frac{\hbar^2}{2m_eR^2}\left(m+\frac{\Phi}{\Phi_0}\right)^2,$$
where $m=0,\pm1,\pm2,\pm3...$ is the electron angular momentum
along the ring axis, and $\Phi=B\pi R^2$ is the magnetic flux
through the ring. In the considered case of a mesoscopic ring, it
is convenient to rewrite this spectrum as
\begin{equation}\label{B}
\varepsilon(k)=\frac{\hbar^2}{2m_e}\left(k+\frac{\Phi}{R\Phi_0}\right)^2,
\end{equation}
where $k=m/R$ is the electron wave vector along the ring.
Graphically, the energy spectrum (\ref{B}) can be pictured as a
parabola shifted along the $k$ axis by the wave vector
\begin{equation}\label{k0}
k_0=-\Phi/R\Phi_0
\end{equation}
(see Fig.~\ref{fig1}b). Formally, just the wave vector (\ref{k0})
defines the nonzero phase shift (\ref{tau}) since $\Delta\phi=2\pi
R k_0$.

Any electromagnetic field, which results to such a shifted
electron energy spectrum with $k_0\neq0$, can generate the
oscillations of conductance of the considered electron
interference device. However, in the case of a time-dependent
electromagnetic field with the vector potential ${A}_\varphi(t)$,
the Schr\"odinger equation with the Hamiltonian (\ref{H0}) is
non-stationary and cannot be used to find the electron energy
spectrum. The regular approach to solve this quantum-mechanical
problem should be based on the conventional methodology of quantum
optics. \cite{Scully,Cohen-Tannoudji} Namely, we have to consider
the system ``electrons in the ring + electromagnetic field'' as a
whole and to write the Hamiltonian of this electron-photon system.
If the field frequency lies far from the resonant frequencies of
the electron subsystem (i.e. the field is purely ``dressing''),
then the energy spectrum of the electron-photon system can be
written as a sum of field energy and energy of the electrons
strongly coupled to the field (dressed electrons). This energy
spectrum of dressed electrons will be responsible for all electron
characteristics of the ring subjected to the strong high-frequency
electromagnetic field.

The Hamiltonian (\ref{H0}) is written as a function on the vector
potential $A_\varphi(t)$ which depends on the gauge. In order to
rewrite the Hamiltonian in gauge invariant form, let us apply the
unitary transformation \cite{Scully}
$$U=\exp\left(\frac{ieR}{\hbar}\int A_\varphi(t)d\varphi\right),$$ where the indefinite integral over the angle $\varphi$ should be
treated as an anti-derivative of the integrand. Then the
transformed electron Hamiltonian (\ref{H0}),
$$\hat{H}^\prime_0=U^\dagger \hat{H}_0U-i\hbar U^\dagger \frac{\partial U}{\partial t},$$
takes the form
\begin{equation}\label{HP}
\hat{H}^\prime_0(E_\varphi)=\frac{\hat{p}^2_\varphi}{2m_e}-eR\int
E_\varphi d\varphi,
\end{equation}
where $E_\varphi=-\partial A_\varphi(t)/\partial t$ is the angular
component of the electric field which does not depend on the field
gauge. Although the interaction of electrons in ringlike
structures with an electric field was considered previously (see,
e.g., Ref.~[\onlinecite{Kibis2005,Portnoi2012}]), phase-shift
phenomena caused by a high-frequency field have so far escaped
attention. Considering the problem within the conventional
quantum-field approach, \cite{OxfordQE,Scully,Cohen-Tannoudji} the
classical electric field in the Hamiltonian (\ref{HP}),
$\mathbf{E}$, should be replaced with the field operator,
$\hat{\mathbf{E}}$. Then the complete electron-photon Hamiltonian
reads
\begin{equation}\label{efH}
\hat{H}=\sum_{\mathbf{q}}\hbar\omega_{\mathbf{q}}
\hat{a}^\dagger_{\mathbf{q}}\hat{a}_{\mathbf{q}}+\hat{H}^\prime_0(\hat{E}_\varphi),
\end{equation}
where the first term describes the field energy, $\mathbf{q}$ is
the photon wave vector, $\omega_{\mathbf{q}}$ is the photon
frequency, $\hat{a}^\dagger_{\mathbf{q}}$ and
$\hat{a}_{\mathbf{q}}$ are the photon operators of creation and
annihilation respectively, and the summation is assumed to be
performed over all photon modes of the electromagnetic field. If
the ring is subjected to a monochromatic circularly polarized
electromagnetic wave propagating perpendicularly to the ring, the
Hamiltonian (\ref{efH}) takes the form
\begin{equation}\label{AH}
\hat{\mathcal
H}=\hbar\omega\hat{a}^\dagger\hat{a}+\frac{\hat{p}_\varphi^2}{2m_e}
-{ieR}\sqrt{\frac{\hbar\omega}{4\epsilon_0V}}\left(e^{i\varphi}\hat{a}-e^{-i\varphi}\hat{a}^\dagger\right),
\end{equation}
where $\omega$ is the field frequency. Considering the last term
in the Hamiltonian (\ref{AH}) as a perturbation, we can apply the
approach developed in Ref.~\onlinecite{Kibis2011} to solve the
electron-photon Schr\"odinger equation with this Hamiltonian. From
experimental viewpoint, the most relevant case corresponds to the
ring exposed to a classically strong laser-generated
electromagnetic field. Just such a strong electromagnetic field
will be under consideration in the following. In contrast to the
case of a ring interacting with a weak photon mode inside a
cavity, \cite{Kibis2013,Arnold2013} an amplitude of the strong
field does not depend on the electron-photon interaction. As a
result, the energy spectrum of dressed electrons in the ring can
be found as an expansion in terms of the dressing field amplitude
$E_0$. Assuming the inequality $|e|E_0/m_eR\omega^2\ll1$ to be
satisfied and accounting for terms squared in the field amplitude
only, the energy spectrum of dressed electrons in the ring can be
written as
\begin{equation}\label{ED}
\varepsilon(k)=\frac{\hbar^2k^2}{2m_e}+\frac{\hbar e^2E_0^2}
{2m_e^2R\omega^3}k.
\end{equation}
It should be noted that the Hamiltonian (\ref{AH}) describes
electrons in an isolated ring, where the electron lifetime is
$\tau\rightarrow\infty$. In the interference device pictured in
Fig.~\ref{fig1}a, this lifetime is the traveling time of an
electron from one QPC to the other one, i.e. $\tau\sim \pi R/v_F$,
where $v_F$ is the Fermi velocity of an electron in the ring.
Therefore, the developed theory is consistent if the field
frequency, $\omega$, is large enough to satisfy the condition
$2\pi/\omega\tau\ll1$ which allows one to consider the incident
electromagnetic field as a dressing field.

The energy spectrum (\ref{ED}) has the form plotted in
Fig.~\ref{fig1}b with
\begin{equation}\label{k00}
k_0=-\frac{e^2E_0^2}{2m_e\hbar R\omega^3}.
\end{equation}
It follows from the comparison of Eqs.~(\ref{k0}) and (\ref{k00})
that the high-frequency electromagnetic field results in the same
phase shift (\ref{tau}) as an effective magnetic flux
\begin{equation}\label{F}
\Phi_{\text{eff}}=\frac{|e| \pi E_0^2}{m_e\omega^3}.
\end{equation}
Let us show that the effective magnetic flux (\ref{F}) can be
described in terms of an artificial $U(1)$ gauge field with the
vector potential
\begin{equation}\label{Aeff}
A_\varphi^{\text{eff}}=\frac{\Phi_{\text{eff}}}{2\pi
R}=\frac{|e|E_0^2}{2Rm_e\omega^3},
\end{equation}
which is produced by the strong electron-photon coupling.

\section{The artificial gauge field} To describe the electron-photon coupling in the considered system,
let us use the joined electron-photon space
$|m,N\rangle=|\psi_m(\varphi)\rangle\otimes|N\rangle$. This
corresponds to the electromagnetic field being in the state with
the photon occupation number $N=1,2,3,...$, and the electron being
in the state with the wave function
\begin{equation}\label{psi}
\psi_m(\varphi)=\sqrt{1/2\pi}e^{im\varphi},
\end{equation}
where $m=0,\pm1,\pm2,...$ is the electron angular momentum along
the ring axis. The electron-photon states $|m,N\rangle$ are true
eigenstates of the Hamiltonian of the noninteracting
electron-photon system,
\begin{equation}\label{H00}
\hat{\mathcal
H}^{(0)}=\hbar\omega\hat{a}^\dagger\hat{a}+\frac{\hat{p}_\varphi^2}{2m_e},
\end{equation}
and their energy spectrum is
\begin{equation}\label{E0}
\varepsilon^{(0)}_{m,N}=N\hbar\omega+\frac{\hbar^2m^2}{2m_eR^2}.
\end{equation}
Considering the last term in the Hamiltonian (\ref{AH}) as a
perturbation with the matrix elements
\begin{eqnarray}\label{UM}
\langle
m^\prime,N^\prime|\hat{U}|m,N\rangle&=&-ieR\sqrt{\frac{\hbar\omega}{4\epsilon_0V}}
\left[\sqrt{N}\delta_{m,m^\prime-1}\delta_{N,N^\prime+1}\right.\nonumber\\
&-&\left.\sqrt{N+1}\delta_{m,m^\prime+1}\delta_{N,N^\prime-1}\right]
\end{eqnarray}
and performing trivial calculations within the first order of the
perturbation theory, we can write eigenstates of the Hamiltonian
(\ref{AH}) as
\begin{eqnarray}\label{ES}
|\Psi_{m,N}\rangle=\frac{\langle
m+1,N-1|\hat{U}|m,N\rangle}{\varepsilon^{(0)}_{m,N}-\varepsilon^{(0)}_{m+1,N-1}}|m+1,N-1\rangle\nonumber\\
+\frac{\langle
m-1,N+1|\hat{U}|m,N\rangle}{\varepsilon^{(0)}_{m,N}-\varepsilon^{(0)}_{m-1,N+1}}|m-1,N+1\rangle+|m,N\rangle.
\end{eqnarray}
Substituting Eqs.~(\ref{E0})--(\ref{UM}) into Eq.~(\ref{ES}) and
assuming the electromagnetic field to be strong ($N\gg1$), we
arrive at the expression
\begin{eqnarray}\label{ES1}
|\Psi_{m,N}\rangle&=&|m,N\rangle-\frac{ieRE_0}{2}\left[\frac{|m+1,N-1\rangle}{\hbar\omega-\varepsilon_R(1+2m)}\right.\nonumber\\
&+&\left.\frac{|m-1,N+1\rangle}{\hbar\omega+\varepsilon_R(1-2m)}\right],
\end{eqnarray}
where $E_0=\sqrt{N\hbar\omega/\epsilon_0V}$ is the classical
amplitude of electric field, and $\varepsilon_R=\hbar^2/2m_eR^2$
is the characteristic electron energy in the ring. Taking into
account Eq.~(\ref{psi}), we can rewrite the basis electron-photon
states as $|m\pm1,N\rangle=e^{\pm i\varphi}|m,N\rangle$. Then the
eigenstates (\ref{ES1}) takes the form
\begin{eqnarray}\label{ES2}
|\Psi_{m,N}\rangle&=&|m,N\rangle-\frac{ieRE_0}{2}\left[\frac{e^{i\varphi}|m,N-1\rangle}{\hbar\omega-\varepsilon_R(1+2m)}\right.\nonumber\\
&+&\left.\frac{e^{-i\varphi}|m,N+1\rangle}{\hbar\omega+\varepsilon_R(1-2m)}\right].
\end{eqnarray}
In the basis of the three electron-photon states,
\begin{equation}\label{b}
\begin{pmatrix}
|m,N+1\rangle \\
|m,N\rangle \\
|m,N-1\rangle
\end{pmatrix}
\end{equation}
the eigenstate (\ref{ES2}) can be written formally as a vector
\begin{equation}\label{M}
|\chi\rangle=
\begin{pmatrix}
-\frac{{ieRE_0}/{2}}{\hbar\omega+\varepsilon_R(1-2m)}e^{-i\varphi}\\
1 \\
-\frac{{ieRE_0}/{2}}{\hbar\omega-\varepsilon_R(1+2m)}e^{i\varphi}
\end{pmatrix}.
\end{equation}
It should be noted that each of the basis states (\ref{b})
corresponds to the same electron angular momentum $m$. Therefore,
the influence of the electromagnetic field on the electron results
only in the phase incursion describing by the exponential factors
$e^{\pm i\varphi}$ in the state vector (\ref{M}). Following the
conventional theory of artificial gauge fields (see, e.g.,
Ref.~\onlinecite{Dalibard2011}), we can introduce the $U(1)$ field
with the vector potential, $\mathbf{A}^{\text{eff}}=(i\hbar/e)
\langle{\chi}|\nabla|\chi\rangle$, which corresponds to this phase
incursion. In the case of the ring, this vector potential has the
form $\mathbf{A}^{\text{eff}}=(0,0,A^{\text{eff}}_\varphi)$, where
\begin{equation}\label{Af}
{A}^{\text{eff}}_\varphi=\frac{i\hbar}{eR}\left\langle\chi\left|\frac{\partial}{\partial\varphi}\right|\chi\right\rangle.
\end{equation}
Substituting Eq.~(\ref{M}) into Eq.~(\ref{Af}), we arrive at the
expression
\begin{eqnarray}\label{A1}
A^{\text{eff}}_\varphi&=&\frac{\hbar
eRE_0^2}{4}\left[\frac{1}{[\hbar\omega+\varepsilon_R(1-2m)]^2}\right.\nonumber\\
&-&\left.\frac{1}{[\hbar\omega-\varepsilon_R(1+2m)]^2}\right].
\end{eqnarray}
Under the condition $\hbar\omega\gg\varepsilon_R$, the artificial
vector potential (\ref{A1}) takes the form (\ref{Aeff}).

\section{Discussion and conclusions} Replacing the magnetic flux $\Phi$ with the pseudo-flux (\ref{F})
in known expressions which describe the oscillations of the
conductance of the considered interference device, we can
calculate them as follows.
\begin{figure}[th]
\includegraphics[width=0.48 \textwidth]{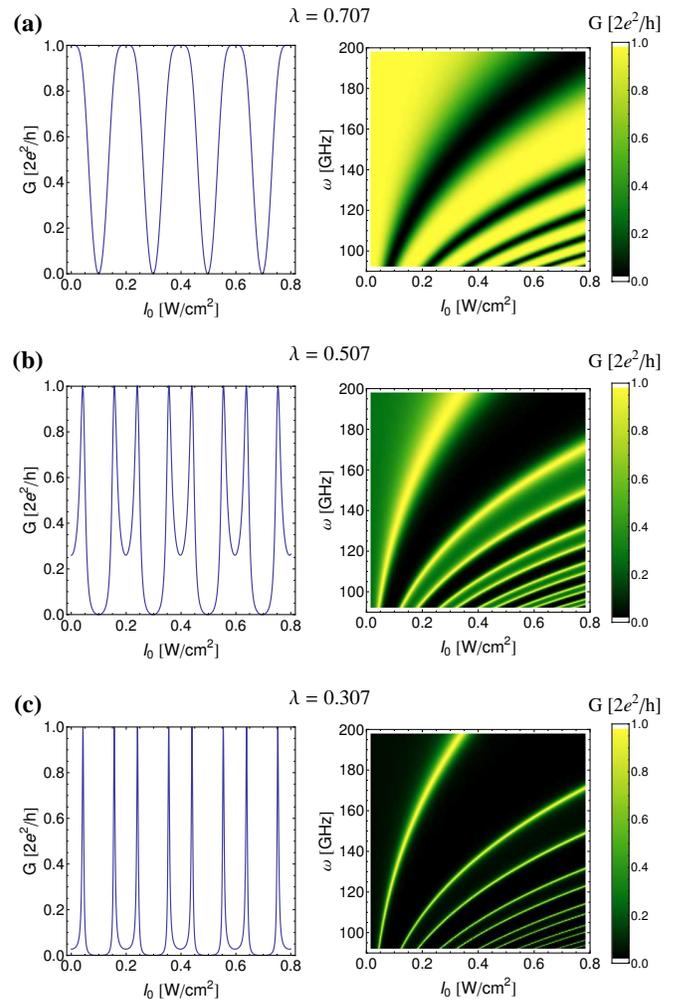}
\caption{(Color online) Conductance of a mesoscopic ring, $G$,
under a circularly polarized electromagnetic wave as a function of
wave intensity $I_0$ and wave frequency $\omega$. Plots (a), (b)
and (c) correspond to different transmission amplitudes $\lambda$
between the current leads and the ring. Frames in the left column
are fixed at the wave frequency $\omega = 100$ GHz. In all plots,
the ring parameters are assumed to be  $R = 10$ $\mu$m,
$\varepsilon_F = 10$ meV, and $m_e = 0.1 \ m_{e0}$, where $m_{e0}$
is the mass of free electron.} \label{fig2}
\end{figure}

First of all, let us consider the ballistic regime. In this case,
the conductance is described by the Landauer formula $G
=(2e^2/h)|C|^2$, where the transmission amplitude of the
interference device, $C$, depends on the coupling between the
leads and the ring. Generally, this coupling can be described by
lead-to-ring and ring-to-lead transmission amplitudes, $\lambda$.
\cite{Gefen1983,Buttiker1984,Shelykh2005} If the reflection from
one lead to itself is absent (i.e., there is no electron
backscattering from QPCs), the transmission amplitude is
$\lambda=\pm1/\sqrt{2}$. This corresponds to the incoming electron
wave being divided equally in the ring along the clockwise
($\phi_+$) and counterclockwise ($\phi_-$) paths (see Fig.~1a). In
this simplest case, the replacement
$\Phi\rightarrow\Phi_{\text{eff}}$ in the expression describing
the AB-oscillations \cite{Shelykh2005} yields
\begin{equation}\label{C}
G =\frac{2e^2}{h}\left[ 1 -
\left|\frac{\sin^2(\Phi_{\text{eff}}/2\Phi_0)}{1-\exp(i2\pi
Rk_F)\cos^2(\Phi_{\text{eff}}/2\Phi_0)}\right|^2\right],
\end{equation}
where $k_F$ is the Fermi electron wave vector in the ring. For
other amplitudes $\lambda$, the conductance $G$ can be calculated
numerically by using the same theory.
\cite{Gefen1983,Buttiker1984,Shelykh2005} Results of the
calculations for different amplitudes $\lambda$ are presented in
Fig.~\ref{fig2}.
\begin{figure}[th]
\includegraphics[width=0.48 \textwidth]{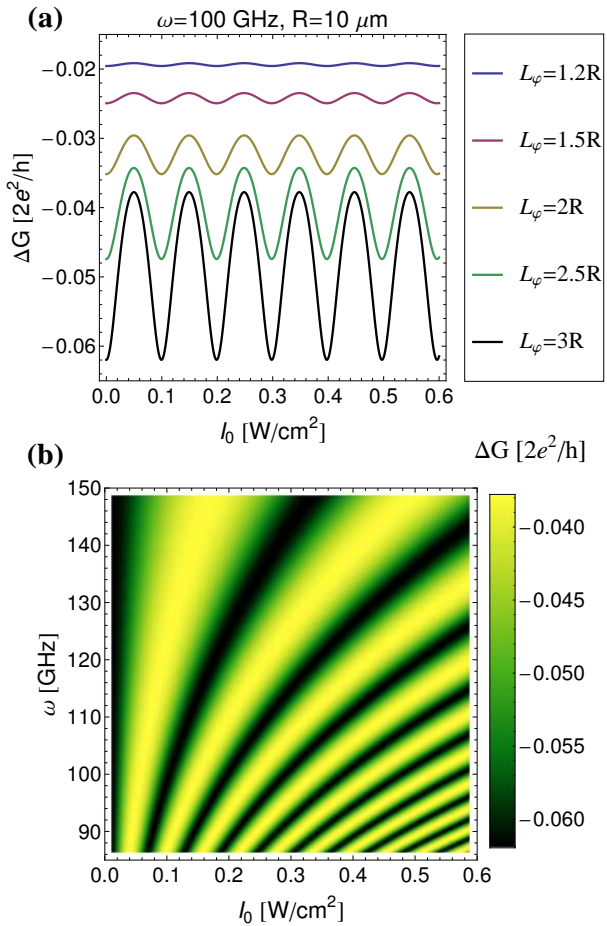}
\caption{(Color online) Weak-localization correction to the
conductance of a mesoscopic ring, $\Delta G$, under a circularly
polarized electromagnetic field: (a) the correction is plotted as
a function of field intensity $I_0$ for different values of
$L_\varphi$ with $\omega = 100$ GHz and $R=10\mu$m; (b) the
correction is plotted as a function of field intensity $I_0$ and
of field frequency $\omega$ for $L_\varphi = 3R$ and $R=10\mu$m.}
\label{fig3}
\end{figure}

For absolutely transparent QPCs ($\lambda = 0.707$), the regular
AB-like oscillations take place (Fig.~\ref{fig2}a). Decreasing the
transparency (decreasing $\lambda$) changes the shape of the
oscillation pattern (Figs.~\ref{fig2}b and \ref{fig2}c). In the
Fourier spectrum of the conductance, the role of the higher
harmonics increases, and eventually these harmonics with a half
period become dominant (see Fig.\ref{fig2}c). Physically,
reduction of the period arises from an increased confinement of
electrons inside the ring, caused by the decrease of transparency
of the QPCs. This leads to an increase of the role of round trips
of an electron inside the ring, which results in the increment of
the effective electron path and, as a consequence, decrease of the
period of the oscillations.

In the diffusive regime, the conductivity of a disordered
ring-shaped conductor with the dephasing length $L_{\varphi}$ can
be described by the expression
\begin{equation}
\Delta \sigma = - \frac{e^2 L_\varphi}{\pi^2 h} \frac{\sinh{\left(
2 \pi R/L_\varphi \right)}}{\cosh{\left( 2 \pi R/L_\varphi
\right)} + \cos{\left( 4 \pi \Phi_{\text{eff}}/\Phi_0 \right)} },
\end{equation}
which is derived from the conventional theory of AAS-oscillations
\cite{AAS} by the replacement $\Phi\rightarrow\Phi_{\text{eff}}$.

The weak-localization correction to the conductance, $\Delta G =
\Delta \sigma/\pi R$, is plotted in Fig.~\ref{fig3} for different
values of the dephasing length $L_\varphi$. As expected, the
correction oscillates with a period which is less then the period
of AB-like oscillations (Fig.~\ref{fig2}a) by a factor of 2. As
for the amplitude of the oscillations, it decays exponentially
when the dephasing length $L_\varphi$ is much smaller than the
distance between the QPCs, $\pi R$. Physically, this decay is
caused by the electron waves loosing their coherence quickly. It
should be noted that an electromagnetic field can cause additional
decoherence of electrons in conducting systems
\cite{Altshuler1981,Altshuler1998,Wenzler2008} and, therefore,
influences on the dephasing length $L_\varphi$. However, the
condition of applicability of dressing field model,
$\omega\tau>>1$, corresponds physically to the absence of energy
exchange between conduction electrons and a dressing field, where
$\tau$ is the characteristic electron relaxation time. Therefore,
there is no heating of electrons by the field under this
condition. As a consequence, the photon-induced breaking of phase
coherence is negligibly small for a dressing field. Plotting the
correction to the conductance, $\Delta G$, in Fig.~\ref{fig3}, we
assumed the field to be high-frequency enough to neglect the phase
decoherence arisen from the field.

Summarizing the aforesaid, we have shown that the interference of
electron waves traveling through a mesoscopic ring exposed to a
circularly polarized electromagnetic field is formally the same as
in a ring subjected to a magnetic flux. As a consequence, the
\emph{optically-induced} Aharonov-Bohm effect appears. This effect
manifests itself in the oscillating dependence of the ring
conductance on the field intensity and field frequency. The
periods of the optically-induced oscillations in the ballistic
regime and the diffusive regime differ from each other by a factor
of 2 in the same manner as periods of the oscillations induced by
a magnetic flux. Therefore, the effect can be described formally
in terms of the artificial $U(1)$ gauge field arisen from the
strong electron-photon coupling.

\begin{acknowledgements}
The work was partially supported by FP7 IRSES projects POLATER,
POLAPHEN and QOCaN, FP7 ITN project NOTEDEV, Tier 1 project
``Polaritons for novel device applications'', and RFBR projects
13-02-90600 and 14-02-00033.
\end{acknowledgements}

\end{document}